\documentclass{JAC2003}

\usepackage{graphicx}
\usepackage{booktabs}
\usepackage{multicol}
\usepackage{amsmath}
\usepackage{url}
\begin{document}
\title{
Model of Flux Trapping in Cooling Down Process\thanks{The work is supported by JSPS Grant-in-Aid for Young Scientists (B) Grant Number 26800157, JSPS Grant-in-Aid for Challenging Exploratory Research Grant Number 26600142, and Photon and Quantum Basic Research Coordinated Development Program from the Ministry of Education, Culture, Sports, Science and Technology, Japan.}
}

\author{
Takayuki Kubo\thanks{kubotaka@post.kek.jp}\\
KEK, High Energy Accelerator Research Organization, Tsukuba, Ibaraki, Japan
}

\maketitle

\begin{abstract}
The flux trapping that occurs in the process of cooling down of the superconducting cavity is studied. 
The critical fields $B_{c2}$ and $B_{c1}$ depend on a position when a material temperature is not uniform. 
In a region with $T\simeq T_c$, 
$B_{c2}$ and $B_{c1}$ are strongly suppressed and can be smaller than the ambient magnetic field, $B_a$. 
A region with $B_{c2}\le B_a$ is normal conducting, 
that with $B_{c1}\le B_a < B_{c2}$ is in the vortex state, 
and that with $B_{c1}> B_a$ is in the Meissner state. 
As a material is cooled down, 
these three domains including the vortex state domain sweep and pass through the material. 
In this process, vortices contained in the vortex state domain are trapped by pinning centers distributing in the material. 
A number of trapped fluxes can be evaluated by using the analogy with the beam-target collision event, 
where beams and a target correspond to pinning centers and the vortex state domain, respectively. 
We find a number of trapped fluxes and thus the residual resistance are proportional to the ambient magnetic field and the inverse of the temperature gradient. 
The obtained formula for the residual resistance is consistent with experimental results. 
The present model focuses on what happens at the phase transition fronts during a cooling down, reveals why and how the residual resistance depends on the temperature gradient, and  naturally explains how the fast cooling works. 
\end{abstract}

%%%%%%%%%%%%%%%%%%%%%%
%%%%%%%%%%%%%%%%%%%%%%
\section{Introduction}
%%%%%%%%%%%%%%%%%%%%%%
%%%%%%%%%%%%%%%%%%%%%%

The surface resistance of the superconducting (SC) radio frequency (RF) cavity consists of the temperature dependent part and the temperature independent part. 
The latter is called the residual resistance, $R_{\rm res}$, and limits the quality factor of  SCRF cavity at $T\ll T_c$.

Magnetic fluxes trapped in a process of cooling down of a cavity degrade $R_{\rm res}$.  Thus decreasing a number of trapped fluxes is necessary for a reduction of $R_{\rm res}$. 
Recent studies show that cooling down conditions affect a number of trapped fluxes~\cite{vogt2013, vogt2015, romanenko2014, romanenko2014APL}. 
In particular, researchers in Fermilab found a fast cooling with a larger temperature gradient leads to a better expulsion of fluxes and thus yields a lower  residual resistance $R_{\rm res}$~\cite{romanenko2014, romanenko2014APL}. 
They achieved an ultra high $Q_0\,(\simeq 2\times 10^{11})$ by the fast cooling method~\cite{romanenko2014APL}.

While many experimental studies on the flux trapping have been conducted, 
not much theoretical progress followed on it. 
In the present paper, we theoretically study the flux trapping that occurs in the process of cooling down by focusing on the dynamics in the vicinity of the phase transition fronts. 
We do not consider effects of the thermal current. 
We show a number of trapped fluxes and thus $R_{\rm res}$ are proportional to the ambient magnetic field and the inverse of the temperature gradient. 
The present model reveals why and how the residual resistance depends on the temperature gradient, and  naturally explains how the fast cooling works.

%%%%%%%%%%%%%%%%%%%%%%
%%%%%%%%%%%%%%%%%%%%%%
\section{Model}
%%%%%%%%%%%%%%%%%%%%%%
%%%%%%%%%%%%%%%%%%%%%%

%
\begin{figure}[tb]
   \begin{center}
   \includegraphics[width=1\linewidth]{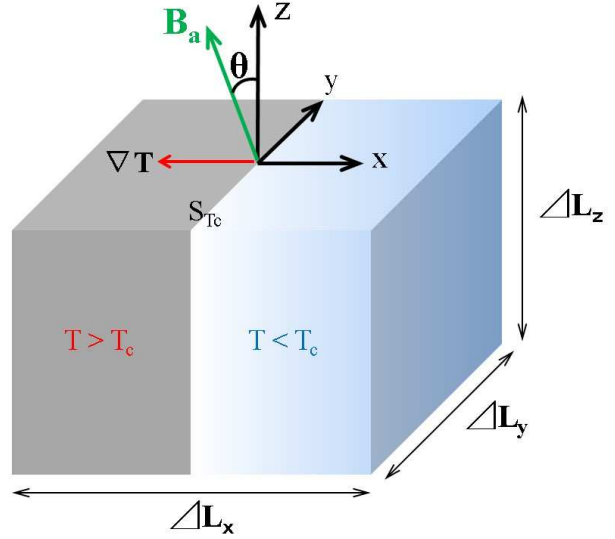}
   \end{center}\vspace{-0.2cm}
   \caption{
SC material of a cavity wall under cooling down. 
The gray and blue region represent the domains with $T>T_c$ and  $T<T_c$, respectively. 
The origin of the $x$-axis is located at the interface of these two regions, 
which we call $S_{T_c}$. 
The SC material is cooled down from the right to the left. 
The ambient magnetic field ${\bf B}_a$ is parallel to the $z$-$x$ plane, 
and $\theta$ is the angle between ${\bf B}_a$ and the $z$-axis. 
   }\label{fig1}
\end{figure}

Let us consider the SC material shown in Figure~\ref{fig1},  
which represents a part of a cavity wall. 
The gray and blue regions represent the domains with $T>T_c$ and  $T<T_c$, respectively. 
The origin of the $x$-axis is located at the interface of these two regions, 
which we call $S_{T_c}$ in the following.   
The SC material is cooled down from the right to the left: 
$\nabla T$ is parallel to the negative direction of the $x$-axis; 
$dT/dx<0$ and $dT/dy=dT/dz=0$. 
The ambient magnetic field ${\bf B}_a$ is parallel to the $z$-$x$ plane, 
and $\theta$ is the angle between ${\bf B}_a$ and the $z$-axis. 
Its magnitude $|{\bf B}_a|$ is given by $B_a$.

\begin{figure}[*t]
   \begin{center}
   \includegraphics[width=1\linewidth]{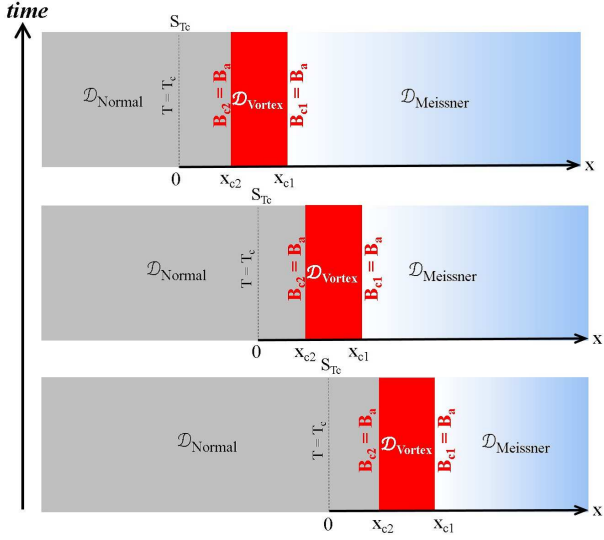}
   \end{center}\vspace{-0.2cm}
   \caption{
Schematic view of the vicinity of the boundary $S_{T_c}$. 
There exist three domains: 
the normal conducting domain $\mathcal{D}_{\rm Normal}$ ($x \le x_{c2}$), 
the vortex state domain $\mathcal{D}_{\rm Vortex}$ ($x_{c2} < x \le x_{c1}$), 
and the Meissner state domain $\mathcal{D}_{\rm Meissner}$ ($x > x_{c1}$). 
As the material is cooled down, the phase transition fronts sweep from the right to the left. 
Vortices contained in $\mathcal{D}_{\rm Vortex}$ are trapped by pinning centers in this process. 
   }\label{fig2}
\end{figure}

Let us look at the vicinity of $S_{T_c}$. 
Since $T \simeq T_c$ in the vicinity of $S_{T_c}$, 
the lower critical field $B_{c1}(T(x))$ and the upper critical field $B_{c2}(T(x))$ are strongly suppressed and can be smaller than the ambient magnetic field $B_a$ (typically $\sim 0.1$-$10 \,\mu{\rm T}$).  
Then we see that there exist two types of the phase transition fronts in the vicinity of $S_{T_c}$: 
$x=x_{c2}$ at which $B_{c2}=B_a$ and $x=x_{c1}$ at which $B_{c1}=B_a$.   
Thus there exist three domains as shown in Fig.~\ref{fig2}: 
\begin{eqnarray}
  \begin{cases}
\mathcal{D}_{\rm Normal}:  & x \le x_{c2}  \\
\mathcal{D}_{\rm Vortex}:  & x_{c2} < x \le x_{c1}  \\
\mathcal{D}_{\rm Meissner}: & x > x_{c1}  
  \end{cases}
\end{eqnarray}
where $\mathcal{D}_{\rm Normal}$, $\mathcal{D}_{\rm Vortex}$, and $\mathcal{D}_{\rm Meissner}$ represent the normal conducting domain, the vortex state domain, and the Meissner state domain, respectively. 
As the SC material is cooled down, 
the domain $\mathcal{D}_{\rm Vortex}$ with the thickness $\delta x \equiv x_{c1}-x_{c2}$ sweeps from the right to the left and finally passes through the entire region of the SC material (see Fig.~\ref{fig2}). 
In this process, vortices contained in $\mathcal{D}_{\rm Vortex}$ are trapped by pinning centers and contribute to $R_{\rm res}$.

%%%%%%%%%%%%%%%%%%%%%%%%%%%%%%%%%%%%%%%%%%%%
%%%%%%%%%%%%%%%%%%%%%%%%%%%%%%%%%%%%%%%%%%%%
\section{Evaluations of the number of trapped fluxes and $R_{\rm res}$}
%%%%%%%%%%%%%%%%%%%%%%%%%%%%%%%%%%%%%%%%%%%%
%%%%%%%%%%%%%%%%%%%%%%%%%%%%%%%%%%%%%%%%%%%%

The goal of this section is to evaluate a number of trapped fluxes and $R_{\rm res}$ based on the model introduced in the last section.   

%%%%%%%%%%%%%%%%%%%%%%%%%%%%%%%%%%%%%%%%%%%%%%%%%%%%
\subsection{Review of the Temperature Dependences of Relevant Parameters}
%%%%%%%%%%%%%%%%%%%%%%%%%%%%%%%%%%%%%%%%%%%%%%%%%%%%

Since we are interested in the vicinity of the phase transition fronts, 
where $T \simeq T_c$, 
we work in the framework of  the Ginzburg-Landau (GL) theory. 
Let us summarize the temperature dependences of relevant parameters in GL theory. 
In the following, we use the normalized temperature, 
\begin{eqnarray}
t \equiv \frac{T}{T_c} \,. 
\end{eqnarray}

According to GL theory, the coherence length is given by
\begin{eqnarray}
\xi(T)
\equiv \xi^* ( 1-t)^{-\frac{1}{2}} 
\label{eq:GLxi}
\end{eqnarray}
with $\xi^* \equiv \sqrt{\gamma \hbar^2/|\alpha_0|}$, 
and the penetration depth is given by
$\lambda(T) = \lambda^* ( 1-t \Bigr)^{-\frac{1}{2}}$ with $\lambda^*=\sqrt{\beta/8\mu_0 e^2 \gamma |\alpha_0|}$, 
where $\alpha_0$, $\beta$, and $\gamma$ are the constants derived from the microscopic theory~\footnote{$\alpha_0$, $\beta$, and $\gamma$ are given by $\alpha_0=-\nu(0)$, $\beta=7\zeta(3)\nu(0)/8\pi^2 k_B^2 T_c^2$, $\gamma=7\zeta(3)\nu(0)v_F^2 /48\pi^2 k_B^2 T_c^2$ for clean SCs, and $\gamma=\pi \nu(0)D/8\hbar k_B T_c$ for dirty SCs, respectively, 
where $\nu(0)$ is the density of state at the Fermi energy, $D$ is a diffusion constant, and $v_F$ is the Fermi velocity.}. 
Thus $\kappa \equiv \lambda(T)/ \xi(T) =\lambda^*/ \xi^*$, 
is a temperature independent parameter in the framework of GL theory. 
The upper critical field is given by
\begin{eqnarray}
&& B_{c2}(T) = \frac{\phi_0}{2\pi \xi(T)^2} = \frac{\phi_0}{2\pi \xi^{*2}} ( 1-t) \,,
\label{eq:GLbc2} 
\end{eqnarray}
and the thermodynamic critical field is given by
$B_{c}(T)=B_{c2}(T)/(\sqrt{2}\kappa) =(\phi_0 /2\sqrt{2}\pi \kappa\xi^{*2}) ( 1-t )$\,, 
where $\phi_0=2.07\times 10^{-15}\,{\rm Wb}$ is the flux quantum. 
When $\kappa \gg 1$, the lower critical field is given by the compact expression, 
\begin{eqnarray}
B_{c1}(T)
= \frac{\ln \kappa + a}{\sqrt{2}\kappa} B_c(T) 
= \frac{\phi_0 (\ln \kappa + a) }{4\pi\kappa^2\xi^{*2}} ( 1-t )\,,
\label{eq:GLbc1}
\end{eqnarray}
where $a \simeq 0.5$. 
In the following, we assume $\kappa \gg 1$ and use Eq.~(\ref{eq:GLbc1}) to evaluate $B_{c1}$. 
This assumption allows us to analytically grasp the physics behind the flux trapping in a cooling down process.

The parameter $\xi^* \equiv \sqrt{\gamma \hbar^2/|\alpha_0|}$, 
which appears in Eqs.~(\ref{eq:GLxi})-(\ref{eq:GLbc1}), 
is related to the coherence length in the microscopic theory, $\xi_{\rm BCS}= \hbar v_F/\pi \Delta(0)$. 
Here we give $\xi^*$ in two special cases as examples:
$\xi^* = 0.74 \xi_{\rm BCS}$  (clean limit) and 
$\xi^* =0.85 \sqrt{\xi_{\rm BCS}\ell}$ (dirty limit), 
where $\ell$ is the electron mean free path.

%%%%%%%%%%%%%%%%%%%%%%%%%%%%%%%%%%%%%%%%%%%%%%%%%%%%
\subsection{Translation into the Position Dependences}
%%%%%%%%%%%%%%%%%%%%%%%%%%%%%%%%%%%%%%%%%%%%%%%%%%%%

%
\begin{figure}[*t]
   \begin{center}
   \includegraphics[width=1\linewidth]{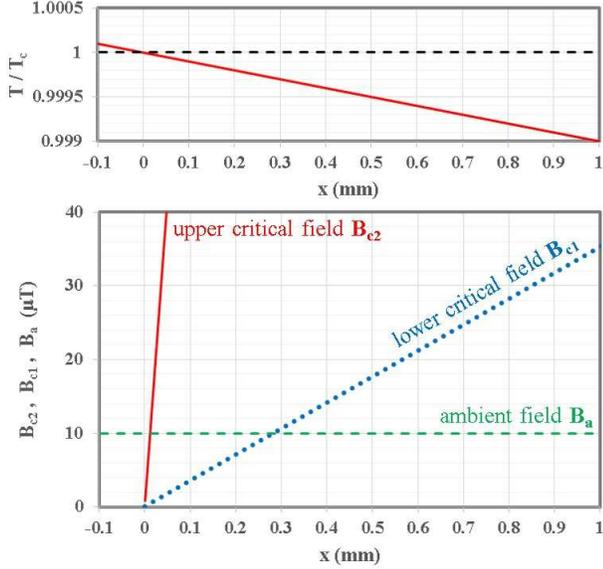}
   \end{center}\vspace{-0.2cm}
   \caption{
Examples of the temperature and the critical field distributions as functions of $x$, 
where $|dt/dx|= T_c^{-1} |dT/dx| = 1\,{\rm m^{-1}}$, $\kappa=5$, and $\xi^* = 20\,{\rm nm}$ are assumed. 
As an example, the ambient magnetic field $B_a = 10\,\mu{\rm T}$ is also shown together. 
The regions $B_a \ge B_{c2}$, $B_{c2}>B_a \ge B_{c1}$, and $B_a < B_{c1}$ correspond to $\mathcal{D}_{\rm Normal}$, $\mathcal{D}_{\rm Vortex}$, and $\mathcal{D}_{\rm Meissner}$, respectively (see also Fig.~\ref{fig2}). 
   }\label{fig3}
\end{figure}

The above parameters as functions of the temperature, $T$, can be translated into functions of the position, $x$, by using the temperature distribution, $T(x)$. 
Since the origin of the $x$-axis is located at the boundary $S_{T_c}$, 
where $T=T_c$, 
the temperature distribution $T(x)$ can be written as  
\begin{eqnarray}
T(x) 
=  T_c + \frac{dT}{dx} x 
= T_c \biggl( 1 - \biggl| \frac{dt}{dx} \biggr| x \biggr) \,. 
\label{eq:Tdistribution}
\end{eqnarray}
By using Eq.~(\ref{eq:Tdistribution}) or $t = 1-|dt/dx| x$, 
we find $1-t = |dt/dx|x$ and $(1-t)^{-\frac{1}{2}}=1/\sqrt{|dt/dx|x}$. 
Then the coherence length given by Eq.~(\ref{eq:GLxi}) can be written as a function of  $x$: 
\begin{eqnarray}
\xi(T(x))=\frac{\xi^*}{\sqrt{x}} \biggl| \frac{dt}{dx} \biggr|^{-\frac{1}{2}} \,. \label{eq:xi_x}
\end{eqnarray}
The critical fields given by Eqs.~(\ref{eq:GLbc2}) and (\ref{eq:GLbc1}) can also be written as functions of $x$: 
\begin{eqnarray}
&&B_{c2}(T(x)) = \frac{\phi_0}{2\pi \xi^{*2}} \biggl| \frac{dt}{dx} \biggr| x \,, 
\label{eq:bc2_x} \\
&&B_{c1}(T(x)) = \frac{\phi_0 (\ln \kappa + a) }{4\pi\kappa^2\xi^{*2}} \biggl| \frac{dt}{dx} \biggr| x \,. 
\label{eq:bc1_x}
\end{eqnarray}
The positions of the phase transition fronts, $x_{c2}$ and $x_{c1}$, 
are given by those at which $B_a$ equals $B_{c2}$ and $B_{c1}$, respectively. 
Substituting $x=x_{c2}$ ($x=x_{c1}$) and $B_{c2}=B_a$ ($B_{c1}=B_a$) into Eq.~(\ref{eq:bc2_x}) (Eq.~(\ref{eq:bc1_x})), we find
\begin{eqnarray}
&&x_{c2} = 2\pi \xi^{*2} \frac{B_a}{\phi_0} \biggl| \frac{dt}{dx} \biggr|^{-1} \,, 
\label{eq:xc2} \\
&&x_{c1} = \frac{4\pi \kappa^2 \xi^{*2}}{\ln \kappa + a} \frac{B_a}{\phi_0} \biggl| \frac{dt}{dx} \biggr|^{-1} \,. 
\label{eq:xc1} 
\end{eqnarray}
Fig.~\ref{fig3} shows $t\equiv T/T_c$, $B_{c2}$ and $B_{c1}$ as functions of $x$. 
The region $B_a \ge B_{c2}$ or $x \le x_{c2}$ corresponds to $\mathcal{D}_{\rm Normal}$,
$B_{c2}>B_a \ge B_{c1}$ or $x_{c2} < x \le  x_{c1}$ corresponds to $\mathcal{D}_{\rm Vortex}$, 
and $B_a < B_{c1}$ or $x> x_{c1}$ corresponds to $\mathcal{D}_{\rm Meissner}$.

It should be noted that all the above calculations are based on the GL theory, 
and are valid near $T\simeq T_c$ or $|dt/dx|x \ll 1$ [see also Eq.~(\ref{eq:Tdistribution})]. 
Thus the positions of the phase transition fronts, $x_{c2}$ and $x_{c1}$, given by Eqs.~(\ref{eq:xc2}) and (\ref{eq:xc1}) are also valid only when $x_{c2}\ll |dt/dx|^{-1}$ and $x_{c1}\ll |dt/dx|^{-1}$, respectively. 
These conditions are satisfied as long as a typical ambient magnetic field $< \mathcal{O}(10)\,\mu{\rm T}$ is assumed. 

%%%%%%%%%%%%%%%%%%%%%%%%%%%%%%%%%%%%%%%%%%%%%%%
\subsection{Number of Trapped Fluxes}
%%%%%%%%%%%%%%%%%%%%%%%%%%%%%%%%%%%%%%%%%%%%%%%

%
\begin{figure}[*t]
   \begin{center}
   \includegraphics[width=1\linewidth]{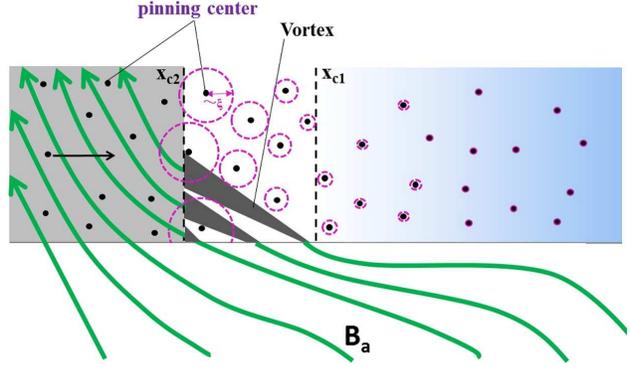}
   \end{center}\vspace{-0.2cm}
   \caption{
On our coordinate system moving with $S_{T_c}$, 
the phase transition fronts, $x=x_{c2}$ and $x=x_{c1}$, are at rest, 
but pinning centers move. 
This can be described as ``beam-target collision" events, 
where beams and a target correspond to pinning centers and the vortex state domain, respectively. 
   }\label{fig4}
\end{figure}

On our coordinate system that moves with the boundary $S_{T_c}$, 
the phase transition fronts are at rest, 
and pinning centers move and collide with vortices (see Fig.~\ref{fig4}). 
Then the flux trapping phenomenon can be described as a beam-target collision event with a  reaction cross-section, $\sigma$,   
where pinning centers and the domain $\mathcal{D}_{\rm vortex}$ correspond to beams and a target, respectively.

We assume the pinning force can reach a distance of the order of the coherence length like that of a grain boundary or a normal conducting precipitate in an SC with $\kappa\gg 1$. 
Note that the size of $\xi$ in the vicinity of the phase transition fronts, where $T\sim T_c$, is much larger than that of $T\ll T_c$ and typically $\mathcal{O}(1) \mu {\rm m}$.  
For example, at the phase transition front $x_{c2}$,  
\begin{eqnarray}
\xi(x_{c2})
= \frac{\xi^*}{\sqrt{x_{c2}}} \biggl| \frac{dt}{dx} \biggr|^{-\frac{1}{2}}_{x=\bar{x}}
= \frac{1}{\sqrt{2\pi}} \sqrt{\frac{\phi_0}{B_a}} 
\,, \label{eq:xiATxc2}
\end{eqnarray}
which yields $\xi(x_{c2})\simeq 6 \mu{\rm m}$ when $B_a = 10\,\mu {\rm T}$, and $\xi(x_{c2})\simeq 60 \mu{\rm m}$ when $B_a = 0.1\,\mu {\rm T}$.

Pinning centers and vortices have effective radii $\sim \xi$. 
Since $\xi$ is maximum at  $x=x_{c2}$ and decreases as $x$ increases,
we assume reactions mostly occur near $x=x_{c2}$ for simplicity (see Fig.~\ref{fig4}). 
Then the reaction cross-section $\sigma$ is given by
\begin{eqnarray}
\sigma \propto  \xi (x_{c2})^2 
\,. \label{eq:sigma}
\end{eqnarray}
where $\xi(x_{c2})$ is given by Eq.~(\ref{eq:xiATxc2}). 

The number of vortices in the ``target" $\mathcal{D}_{\rm vortex}$ is given by 
\begin{eqnarray}
N_{\phi}^{(\delta x)} \propto \frac{B_a}{\phi_0} \Delta L_y \delta x 
\,, \label{eq:Nphideltax}
\end{eqnarray}
where $B_a/\phi_0$ is a vortex density, 
and $\delta x$ is the thickness of domain $\mathcal{D}_{\rm vortex}$ given by  
\begin{eqnarray}
\delta x 
\equiv x_{c1}-x_{c2} 
= 4\pi \xi^{*2} f_n (\kappa)
\frac{B_a}{\phi_0} \biggl| \frac{dt}{dx} \biggr|^{-1} 
\,,  \label{eq:deltax} 
\end{eqnarray}
and $f_n (\kappa) \equiv \kappa^2/(\ln\kappa +a) - 1/2$. 
It should be noted that, 
even if some vortices in $\mathcal{D}_{\rm vortex}$ are trapped in the process of passing through the material, 
$N_{\phi}^{(\delta x)}$ does not decrease, 
because vortices are supplied from the phase transition front at $x=x_{c2}$ as long as the material is in the ambient magnetic field. 
Here we give examples of the typical size of $\delta x$. 
Assuming $\kappa=5$, $\xi^*=20\,{\rm nm}$ and $B_a=10\,\mu{\rm T}$, 
$\delta x=3\times 10^2\,\mu{\rm m}$ for $|dt/dx|=1\,{\rm m^{-1}}$ (see also Fig.~\ref{fig3}) and 
$\delta x=30\,\mu{\rm m}$ for $|dt/dx|=10\,{\rm m^{-1}}$. 
As $|dt/dx|$ increases, the target thickness $\delta x$ and thus $N_{\phi}^{(\delta x)}$ decrease.

Now we can evaluate the number of trapped fluxes. 
Introducing the total cross-section $\Sigma=\sigma N_{\phi}^{(\delta x)}$,  
the reaction probability is given by $P=\Sigma/\Delta L_y \Delta L_z$. 
Then the number of events is given by 
$N_{\rm event} = (\rho_{\rm pin} \Delta L_x \Delta L_y \Delta L_z) P = \rho_{\rm pin}\Delta L_x \sigma N_{\phi}^{(\delta x)}$, 
where $\rho_{\rm pin}$ is the density of pinning centers that have strong enough pinning forces to pin vortices against the forces due to the free energy gradient that prevent the vortex penetration. 
Since the number of trapped flux $N_{\rm trap}$ is expected to be proportional to $N_{\rm event}$, we obtain 
\begin{eqnarray}
N_{\rm trap} 
\propto \rho_{\rm pin} \Delta L_x \sigma N_{\phi}^{(\delta x)} 
\propto B_a \biggl| \frac{dt}{dx} \biggr|^{-1} 
\,. \label{eq:Ntrap}
\end{eqnarray}
The factor $|dt/dx|^{-1}$ comes from the fact that 
the total reaction cross-section $\Sigma$ and thus the reaction probability $P$ is proportional to the thickness of the vortex state domain, $\delta x \propto |dt/dx|^{-1}$.

%%%%%%%%%%%%%%%%%%%%%%
\subsection{Residual Resistance}
%%%%%%%%%%%%%%%%%%%%%%

Let us start from the well-known formula for the residual resistance, $R_{\rm res}^{\rm 100\%} = R_n (B_a/B_{c2})$. 
This formula assumes that 100$\%$ of the ambient magnetic field is trapped. 
$R_{\rm res}^{\rm 100\%}$ is reduced to $R_n$ when $B_a=B_{c2}$. 
This formula can be naturally generalized to $R_{\rm res} = R_n (r_{\rm trap} B_a/B_{c2})$, 
where $r_{\rm trap}$ is the ratio of a number of trapped fluxes $N_{\rm trap}$ to a number of total ambient fluxes, $N_{\phi}^{(\Delta L_x)}$. 
Note that $R_{\rm res}$ is reduced to $R_{\rm res}^{\rm 100\%}$ when $r_{\rm trap}=1$. 
Since the total ambient fluxes is given by $N_{\phi}^{(\Delta L_x)} = (\Delta L_x / \delta x) N_{\phi}^{(\delta x)}$, we obtain $r_{\rm trap} =N_{\rm trap}/N_{\phi}^{( \Delta L_x)} \propto \rho_{\rm pin} \sigma \delta x$ or $r_{\rm trap} = M | dt/dx|^{-1}$, where $M\propto \rho_{\rm pin} \xi^{*2} f_n(\kappa)$ is a material dependent parameter. 
Then we find
\begin{eqnarray}
R_{\rm res} 
= R_n \frac{r_{\rm trap} B_a}{B_{c2}} 
= C B_a \biggl| \frac{dt}{dx} \biggr|^{-1} \,,  \label{eq:Rres}
\end{eqnarray}
where $C  \equiv (R_n/B_{c2}) M \propto  (R_n/B_{c2}) \rho_{\rm pin} \xi^{*2} f_n(\kappa)$.

Put briefly, 
trapped fluxes have the normal cores, 
and the total normal conducting area increases as $N_{\rm trap}$ increases (see also  Ref.~\cite{hasan}): 
\begin{eqnarray}
R_{\rm res} \propto N_{\rm trap} \propto B_a \biggl| \frac{dt}{dx} \biggr|^{-1} \,. \label{eq:Rres_propto}
\end{eqnarray}
As mentioned above, 
the factor $|dt/dx|^{-1}$ comes from the thickness of the vortex state domain $\delta x$. 
As $|dT/dx|$ increases,  $\delta x$ decreases, $\Sigma$ and $P$ decreases, 
$N_{\rm trap}$ decreases, and $R_{\rm res}$ decreases.

It should be noted that even if we use the penetration depth instead of the coherence length as a distance that the pinning force can reach, 
the resultant functional form of $R_{\rm res}$ is unchanged because the penetration depth has the same temperature dependence ($\propto (1-t)^{-\frac{1}{2}}$) as the coherence length.

%%%%%%%%%%%%%%%%%%%%%%%%%%%%%%%%%
\subsection{Range in Application of the Present Model}
%%%%%%%%%%%%%%%%%%%%%%%%%%%%%%%%%

When $|dT/dx|$ is so large that the thickness of the vortex state domain $\delta x$ is smaller than $\xi(x_{c2})$, the present model ceases to be valid. 
Such $|dT/dx|$ is obtained by solving $\xi(x_{c2})=\delta x$ and is given by
\begin{eqnarray}
\biggl| \frac{dt}{dx} \biggr| 
= 4\pi \sqrt{2 \pi} \xi^{*2} f_n(\kappa) \biggl( \frac{B_a}{\phi_0} \biggr) \,. 
\end{eqnarray}
On the other hand, when $|dT/dx|$ is too small, the unitarity is broken. 
Thus a cutoff exists at small $|dT/dx|$, 
where 100$\%$ of the ambient magnetic field is trapped: $R_{\rm res}=R_{\rm res}^{\rm 100\%}$. 

%%%%%%%%%%%%%%%%%%%%%%%%%%%%%%%%%
%%%%%%%%%%%%%%%%%%%%%%%%%%%%%%%%%
%%%%%%%%%%%%%%%%%%%%%%%%%%%%%%%%%
\section{Comparison with Experiments}
%%%%%%%%%%%%%%%%%%%%%%%%%%%%%%%%%
%%%%%%%%%%%%%%%%%%%%%%%%%%%%%%%%%
%%%%%%%%%%%%%%%%%%%%%%%%%%%%%%%%%

%
\begin{figure}[t]
   \begin{center}
   \includegraphics[width=1\linewidth]{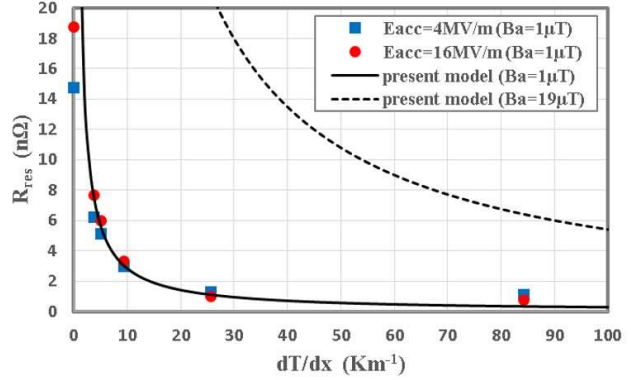}
   \end{center}\vspace{-0.2cm}
   \caption{
$R_{\rm res}$ as functions of $dT/dx$. 
The blue and red symbols represent experimental results read from Ref.~\cite{romanenko2014APL}, where a distance between the two sensors at the equator and the iris is assumed to be $0.1\,{\rm m}$. 
The solid and dashed curves represent Eq.~(\ref{eq:RresExp}) with $B_a =1\,\mu{\rm T}$ and $19\,\mu{\rm T}$, respectively. 
   }\label{fig5}
\end{figure}

We derived $R_{\rm res}$ given by Eq.~(\ref{eq:Rres}) in the last section. 
Now, we compare Eq.~(\ref{eq:Rres}) with experiments by Romanenko et al.~\cite{romanenko2014APL}.

Let us extract the constant $C$ from the experimental results~\cite{romanenko2014APL}. 
Assuming the distance between the temperature sensors at the equator and the iris is given by $\sim 0.1\,{\rm m}$, 
the temperature differences between the equator and iris given in Ref.~\cite{romanenko2014APL} can be translated into the temperature gradients. 
Then $C$ can be evaluated by $C = |dt/dx| R_{\rm res}/B_a$, 
and we find 
\begin{eqnarray}
R_{\rm res} 
=\biggl( \frac{3.08\,{\rm n\Omega}}{{\rm m}\cdot {\rm \mu T}} \biggr)\, 
B_a \biggl| \frac{dt}{dx} \biggr|^{-1} \,, \label{eq:RresExp}
\end{eqnarray}

Fig.~\ref{fig5} shows $R_{\rm res}$ as functions of $|dT/dx|$ ($=T_c |dt/dx|$). 
The blue and red symbols represent experimental results read from Ref.~\cite{romanenko2014APL}, where the temperature differences are translated into the temperature gradients by assuming the distance between the temperature sensors at the equator and iris is given by $\sim 0.1\,{\rm m}$. 
The solid curve represents Eq.~(\ref{eq:RresExp}) with $B_a=1\,\mu {\rm T}\, (10\,{\rm mG})$, 
which agrees well with the experimental results. 
The dashed curve represents Eq.~(\ref{eq:RresExp}) with $B_a=19\,\mu {\rm T}\, (190\,{\rm mG})$. 
We see that, even under such a strong ambient magnetic field, 
$R_{\rm res}\simeq 5\,{\rm n\Omega}$ can be achieved by a large $|dT/dx|\simeq 100\,{\rm K/m}$. 
This result is also consistent with Ref.~\cite{romanenko2014APL}, 
where they achieved $Q_0 > 5 \times 10^{10}$ with the cavity cooled down under $B_a=19\,\mu {\rm T}\, (190\,{\rm mG})$.

Other examples and more detailed discussions will be presented elsewhere~\cite{kubo}.

%%%%%%%%%%%%%%%%%%%%%%%%%%%%%%%%%
%%%%%%%%%%%%%%%%%%%%%%%%%%%%%%%%%
%%%%%%%%%%%%%%%%%%%%%%%%%%%%%%%%%
\section{Summary}
%%%%%%%%%%%%%%%%%%%%%%%%%%%%%%%%%
%%%%%%%%%%%%%%%%%%%%%%%%%%%%%%%%%
%%%%%%%%%%%%%%%%%%%%%%%%%%%%%%%%%

In the present paper, we studied the flux trapping that occurs in the process of cooling down of a superconductor by focusing on the dynamics in the vicinity of the phase transition fronts. 
\begin{itemize}
\item We considered the simple model shown in Fig.~\ref{fig1}. 
The gray and blue region represent the domains with $T>T_c$ and  $T<T_c$, respectively. 
The origin of the $x$-axis is located at the interface of these two regions, 
which we call $S_{T_c}$. 
The SC material is cooled down from the right to the left. 
\item In the vicinity of $S_{T_c}$, where $T \simeq T_c$, 
$B_{c1}$ and $B_{c2}$ are strongly suppressed and can be smaller than the ambient magnetic field $B_a \sim 0.1$-$10 \,\mu{\rm T}$.  
There exist two phase transition fronts in the vicinity of $S_{T_c}$: 
$x=x_{c2}$ at which $B_{c2}=B_a$ and $x=x_{c1}$ at which $B_{c1}=B_a$. 
Then there exist three domains: 
the normal conducting domain ($x \le x_{c2}$), 
the vortex state domain ($x_{c2} < x \le x_{c1}$), 
and the Meissner state domain ($x > x_{c1}$).  
See Fig.~\ref{fig2} and \ref{fig3}. 
\item As the material is cooled down, 
the phase transition fronts together with the vortex state domain sweep the material. 
Vortices contained in the vortex state domain are trapped by pinning centers in this process (see Fig.~\ref{fig2}). 
\item A number of trapped fluxes $N_{\rm trap}$ were evaluated by using the analogy with the beam-target collision event. Beams and a target correspond to pinning centers and vortex state domain, respectively (see Fig.~\ref{fig4}). 
\item The number of trapped fluxes $N_{\rm trap}$ is given by Eq.~(\ref{eq:Ntrap}), 
which is proportional to $B_a$ and $|dT/dx|^{-1}$. 
\item Trapped fluxes have normal cores, and the total normal conducting area increases as $N_{\rm trap}$ increases. 
Thus $R_{\rm res}$ is proportional to $N_{\rm trap}$ and is given by Eq.~(\ref{eq:Rres}),  
which is also proportional to $B_a$ and $|dT/dx|^{-1}$. 
\item The factor $|dT/dx|^{-1}$ in $N_{\rm trap}$ and $R_{\rm res}$ comes from the fact that the thickness of the vortex domain is proportional to $|dT/dx|^{-1}$: 
As $|dT/dx|$ increases, 
the thickness of the vortex state domain decreases, 
the total reaction cross-section and the reaction probability decrease, 
then a number of trapped fluxes $N_{\rm trap}$ decreases, 
and the residual resistance, $R_{\rm res}$ decreases. 
\item  The residual resistance, $R_{\rm res}$, given by Eq.~(\ref{eq:Rres}) was compared with experimental results (see Fig.~\ref{fig5}). 
The blue and red symbols represent experimental results read from Ref.~\cite{romanenko2014APL}, where the temperature differences in the original paper are translated into the temperature gradients. 
The solid curve represents Eq.~(\ref{eq:RresExp}) with $B_a=1\,\mu {\rm T}$, 
which agrees well with the experimental results. 
Note that our formula for $R_{\rm res}$ contains only one free parameter. 
\item The dashed curve in Fig.~\ref{fig5} represents Eq.~(\ref{eq:RresExp}) with $B_a=19\,\mu {\rm T}\, (190\,{\rm mG})$. 
We see that, even under such a strong ambient magnetic field, 
$R_{\rm res}\simeq 5\,{\rm n\Omega}$ can be achieved by a large $|dT/dx|\simeq 100\,{\rm K/m}$. 
This result is also consistent with Ref.~\cite{romanenko2014APL}, 
where they achieved $Q_0 > 5 \times 10^{10}$ with the cavity cooled down under $B_a=19\,\mu {\rm T}\, (190\,{\rm mG})$. 
\end{itemize}

The present model revealed why and how $R_{\rm res}$ depends on the temperature gradient, and  naturally explained how the fast cooling works. 
Other examples and more detailed discussions will be presented elsewhere~\cite{kubo}.


\begin{thebibliography}{99}

\bibitem{vogt2013}
J. M. Vogt et al., 
Phys. Rev. ST Accel. Beams {\bf 16}, 102002 (2013). 

\bibitem{vogt2015}
J. M. Vogt et al., 
Phys. Rev. ST Accel. Beams {\bf 18}, 042001 (2015). 

\bibitem{romanenko2014}
A. Romanenko et al., 
J. Appl. Phys {\bf 115}, 184903 (2014). 

\bibitem{romanenko2014APL}
A. Romanenko et al., 
Appl. Phys. Lett. {\bf 105}, 234103 (2014).

\bibitem{hasan}
H. Padamsee, J. Knobloch, and T. Hays, {\it RF Superconductivity for Accelerators} (John Wiley, New York, 1998). 

\bibitem{kubo}
T. Kubo, to be presented (2015).

\end{thebibliography}
\end{document}